\documentclass{svjour3}         
\usepackage{graphicx}
\usepackage{amssymb}

\begin{document}

\title{Half-open Penning trap with efficient light collection for precision laser spectroscopy of highly charged ions}



\author{David von Lindenfels \and
        Manuel Vogel         \and
        Wolfgang Quint   \and
        Gerhard Birkl \and
        Marco Wiesel
}


\institute{
David von Lindenfels 
\at GSI Helmholtz-Zentrum f\"ur Schwerionenforschung, Planckstrasse 1, D-64291 Darmstadt
\and
Manuel Vogel 
\at Institut f\"ur Angewandte Physik, Technische Universit\"at Darmstadt, D-64289 Darmstadt,\\ e-mail: m.vogel@gsi.de
\and
Wolfgang Quint 
\at
GSI Helmholtz-Zentrum f\"ur Schwerionenforschung, Planckstrasse 1, D-64291 Darmstadt
\and
Gerhard Birkl 
\at Institut f\"ur Angewandte Physik, Technische Universit\"at Darmstadt, D-64289 Darmstadt
\and
Marco Wiesel
\at
 Institut f\"ur Angewandte Physik, Technische Universit\"at Darmstadt, D-64289 Darmstadt
}

\date{Received: date / Accepted: date}

\maketitle

\begin{abstract}
We have conceived, built and operated a 'half-open' cylindrical Penning trap for the confinement and laser spectroscopy of highly charged ions. This trap
allows fluorescence detection employing a solid angle which is about one order of magnitude larger than in conventional cylindrical Penning traps. At the same time, the desired electrostatic and magnetostatic properties of a closed-endcap cylindrical Penning trap are preserved in this configuration. We give a detailed account on the design and confinement properties, a characterization of the trap and show first results of light collection with in-trap produced highly charged ions.

\end{abstract}

\section{Introduction}
Optical spectroscopy of ions in Penning traps \cite{MAJ04,werth09} is naturally hampered by the electrode configuration surrounding the confinement region from which the desired light is emitted. This is particularly true for hyperbolic Penning traps which have good confinement properties by design, but are nearly completely intransparent \cite{gho}. A common solution is to make use of artificial openings which however affect the confining potential and hence need to be kept small. The cylindrical Penning trap with open endcaps as presented by Gabrielse et al. \cite{gab89} solves this problem to some extent while maintaining well-defined confinement conditions close to the trap centre. In this design, the light collection efficiency is limited by the required ratio of diameter to length of the arrangement. Radial trap openings as used in experiments like \cite{bhar,spec,spec2} circumvent this limitation, but are likely to be unsuitable when high precision of the confining fields comparable to experiments like \cite{haff,verd,sturmSi,wag} is required. Also, they necessitate either complicated light collection schemes \cite{bhar} or radial access to the trap \cite{spec,spec2}. Open access to confined particles has been a good part of the motivation when designing planar Penning traps \cite{pla,pla2,gal,bush,gabbash} as well as numerous other types of traps such as wire Penning traps \cite{cast}, stylus traps \cite{wine} and others \cite{a1,a2,a3}. In such a configuration light collection can be immensely effective and closely approach a $4\pi$ open detection angle. However, such traps unfortunately are unsuited for precision spectroscopy of large ensembles of ions as presently discussed \cite{pra1}. Hence, we have devised a specific modification to a cylindrical closed-endcap Penning trap which in the following will be presented in detail. The experiment featuring this trap (ARTEMIS) aims at high precision measurements of electronic and nuclear magnetic moments by laser-microwave double-resonance spectroscopy \cite{pra1} of confined highly-charged ions. A scientific motivation and further details have been given in \cite{pra1,dr,zeeman,pr}. 

\section{Trap Design Considerations}\label{sec:design}
We will briefly discuss the electromagnetic (confinement) properties of cylindrical Penning traps with different choice of geometry to motivate the actual trap design.

\subsection{Open-Endcap Cylindrical Penning Trap}
The concept of approximating the harmonic potential of a hyperbolic Penning trap with an arrangement of three or five cylindrical electrodes with axially open endcaps has been described in detail by Gabrielse et al. (see for example \cite{gab89,gab}). Without loss of generality, a central ring electrode is kept on a constant voltage $U_0$ while the two endcaps are grounded. Radial confinement is provided by the homogeneous magnetic field in which the trap is immersed. If three electrodes are used, the axial and radial dimensions of the electrodes have to fulfill certain criteria ('mechanical compensation') in order for the confining potential to be harmonic around the trap centre \cite{gab}. As this imposes close bounds to the operation parameters, a common choice is the use of two additional electrodes ('compensation electrodes') placed on either side between the ring and the endcaps, and to apply a certain compensation voltage $U_c$ to those. For most applications, it is desired to confine the particles in a harmonic potential such that their oscillation frequencies are well-defined (predictable) and independent of the motional energy, which is presently of importance for non-destructive ion detection and resistive cooling of the ion motion.  

\begin{figure}[h]
\includegraphics[width=\textwidth]{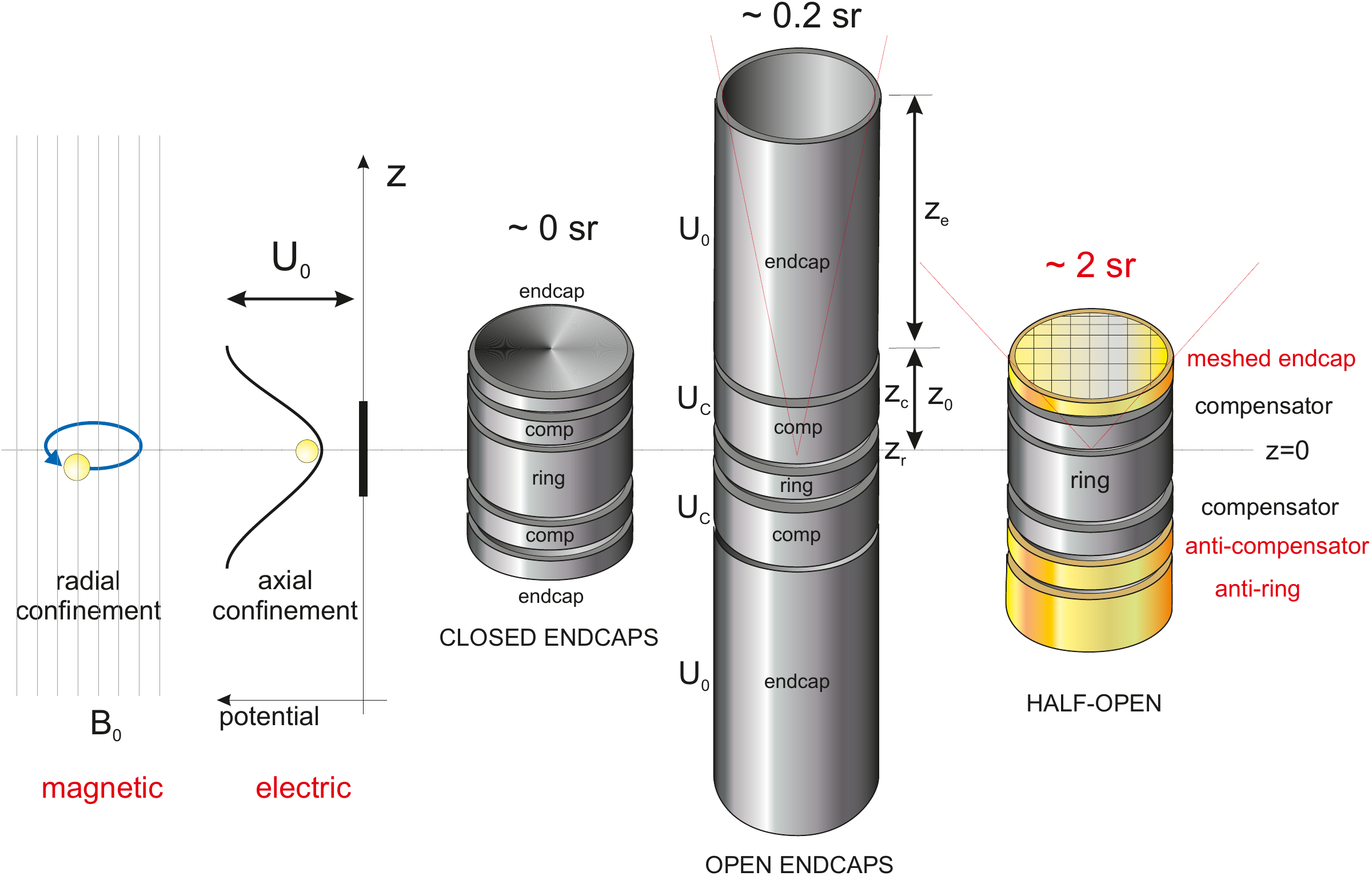}
\caption{Schematic of different types of cylindrical Penning traps and illustration of ion confinement principles. The dashed horizontal line indicates $z=0$ on which the trap centres are located. For details see text}
\label{trap}
\end{figure}

Figure \ref{trap} schematically shows different kinds of cylindrical Penning traps and illustrates the working principle: radial ion confinement is achieved by a constant and homogeneous magnetic field $B_0$ and a superimposed harmonic electric potential created by voltages applied to trap electrodes. The left hand trap is an electrically compensated (5-pole) trap with closed endcaps, as e.g. used in the free electron $g$-2-experiments by Gabrielse et al. \cite{han}.
The middle trap is an electrically compensated (5-pole) trap with open endcaps as used for example in the bound electron magnetic moment measurements \cite{haff,verd,sturmSi,wag}. 
The inner diameter of the cylinders is $2\rho_0$, the axial trap size is characterized by the distance $z_0$ from the trap centre at $z=0$ to one endcap (which for the 5-pole trap is given by $z_0=z_r/2+2z_g+z_c$, i.e. half of the ring length $z_r$ plus two gap sizes $z_g$ between electrodes plus the length $z_c$ of the compensation electrode, see also figure \ref{trap}).
The solid angle of light collection from the trap centre is indicated by dashed lines. The trap on the right-hand side is a modification ('half-open' trap) which allows more efficient light collection from above and open access for ion injection from below. We will now discuss the requirements for harmonic confinement in such a trap.

Generally, we can write the electrostatic potential near the trap centre by the expansion  
\begin{equation}
\label{ela}
\Phi=U_0 \sum_{k=0}^{\infty} c_k \left( \frac{r}{d} \right)^k P_k(\cos \theta) \;\;\;\; \mbox{with} \;\;\;\; d^2=\frac{1}{4} \rho_0^2+\frac{1}{2}z_0^2
\end{equation} 
where $r$ is the radial distance to the trap centre and $P_k(\cos \theta)$ are Legendre polynomials of the $k$-th order with the argument $\cos(\theta)=z/r$, $z$ being the axial distance to the trap centre.  
Ideally, only the coefficients $c_0$ and $c_2$ have non-zero values.  
$c_0$ is an overall potential offset and hence irrelevant for the ion motion. $c_2$  is the quadrupole term creating a harmonic potential well from the applied voltage $U_0$. Using the definition (\ref{ela}), a hyperbolic Penning trap has $c_2=2$ by design. Closed-endcap cylindrical Penning traps have $c_2 \approx 2$, while open-endcap configurations have $c_2 \approx 1$. In cylindrical Penning traps like the present ones, the dominant
electric imperfection is characterized by the term $c_4$. In a cylindrical trap with compensation electrodes, $c_2$ and $c_4$ (and higher-order terms) depend also on applied compensation voltage $U_c$ and can be written as \cite{gab89}
\begin{equation}
\label{tune}
c_2=e_2+d_2 \frac{U_c}{U_0} \hspace{0.5cm} \mbox{and} \hspace{0.5cm} c_4=e_4+d_4 \frac{U_c}{U_0} 
\end{equation}
where $e_k$ and $d_k$ are given by the trap geometry \cite{gab89}. It is thereby possible to minimize the effect of certain imperfections by appropriate choice of
$U_c/U_0$, the so-called tuning ratio ('compensated trap'). The trap is additionally 'orthogonal' if the oscillation frequencies are independent of the applied tuning ratio. The requirements and possibilities for this are discussed in detail in \cite{gab89}.

Only in a harmonic potential, the oscillation frequencies are independent of the kinetic energies of the ion.
For instance, in a potential with a fourth-order perturbation ($c_4 \neq 0$),
\begin{equation}
\Phi(z) = U_0 \left[ c_2 \left( \frac zd \right)^2 +c_4 \left( \frac zd \right)^4 \right], \nonumber
\end{equation}
the finite axial energy $E_z$ causes a relative frequency shift \cite{gab89,sensors} by
\begin{equation}
\frac{\Delta\omega_z}{\omega_z} = \frac{3c_4}{2c_2} \frac{E_z}{M\omega_z^2d^2}
	\label{eq:design-p4}
\end{equation}
A sixth-order perturbation $c_6 (z/d)^6$ can be calculated from equation (\ref{eq:design-p4}) when an effective fourth-order term
\begin{equation}
\tilde{c}_4 = c_4 +\frac{5}{4} c_6  \frac{E_z}{M\omega_z^2d^2} \label{eq:design-p6}
\end{equation}
is used instead of $c_4$ \cite{gab89}. Such energy-dependent frequency shifts need to be minimized when resistive cooling of the ion motion and non-destructive detection are desired to be effective. Also, since the motional
frequencies are used to determine the magnetic field strength (which is an essential requirement for the present application), a high degree of harmonicity of the trapping potential is required. Hence, the coefficients $c_4$ and $c_6$ need to be minimized, which requires correct choice of voltages {\bf and} trap geometry, as will be discussed below for the present trap. 

A drawback of an open-endcap cylindrical design is related to the light collection properties. The fluorescence light of interest emitted from the trap centre has an angular distribution given by
\begin{equation}
I(\theta){\rm d}\theta \propto (1+\cos^2\theta) {\rm d}\theta, \label{eq:distr}
\end{equation}
where $\theta$ is the spherical polar angle. This directional
characteristic prefers emission along the
axis of the magnetic field, which is helpful. However, as the endcap electrodes of open-endcap cylindrical traps need to be very long ($z_e$ at least about $4z_0$ while $z_0 \approx \rho_0$ \cite{gab89}), this leaves
only a small solid angle corresponding to about 1/50 of $4\pi$. 

Before we discuss our solution of choice, it is necessary to briefly discuss the properties of a cylindrical Penning trap with closed endcaps, as we will exploit several of its characteristics and formalisms for potential calculations.

\subsection{Closed-Endcap Cylindrical Penning Trap}
The closed-endcap cylindrical Penning trap (figure \ref{trap}, left) consists of a ring electrode and two end caps,
similar to the hyperbolic trap (a detailed discussion of this can be found in \cite{gab}). Again, a voltage $U_0$ is applied between the ring and the endcaps. Different from the hyperbolic geometry, the ring is a hollow cylinder with a radius
of $\rho_0$ and length $z_r$, and the end caps are flat, having a distance of $z_0$ from the trap
centre. Two correction electrodes (or compensators, used synonymously) next to the ring have a
length $z_c$ and are set to the voltage $U_c$ with respect to the endcaps. The axial spacing between electrodes used as insulation has the amount $z_g$.
The electrostatic potential in the trap volume is a linear superposition of the potentials which are
created by the single electrodes:
\begin{equation}
\Phi(\rho,z) = \sum_{i} U_i \phi^{(i)}(\rho,z). \label{eq:design-superpos}
\end{equation}
The functions $\phi^{(i)}$ are solutions of the Laplace equation with the boundary
conditions $\phi^{(i)}=1$ at the active electrode $i$ and $\phi^{(i)}=0$ at all other electrodes. To
complete the conditions, we assume a linear slope in the gaps between active electrode and its
neighbours, thus forming a trapezoid function. A solution to the Laplace equation can be written as
a linear combination of the basis functions
\begin{equation}
b_n(\rho,z) = \frac{J_0\left(\frac{n \pi \rho}{z_t} \right)}{J_0\left(\frac{n \pi \rho_0}{z_t}\right)}
	\sin\left(\frac{n \pi z}{z_t} \right), \nonumber
\end{equation}
if it is rotation-invariant and zero at the end points $z=0$ and $z=z_t$ \cite{gab}. We
normalize the Bessel function $J_0$ to one at the lateral area of the hollow cylinder ($\rho=\rho_0$).
On this surface, the linear combination $\phi^{(i)} = \sum_n a_n^{(i)} b_n$ can be evaluated,
\begin{equation}
\phi^{(i)}(\rho_0,z) = \sum_n a_n^{(i)} \sin\left(\frac{n \pi z}{z_t} \right),
	\label{eq:design-boundary}
\end{equation}
keeping in mind that these values are specified by the boundary conditions. At the same
time, expression (\ref{eq:design-boundary}) has the form of a Fourier sum. For an electrode
between $z=a$ and $z=b$ (more precisely: with gaps symmetric about $z=a$ and $z=b$) at unit voltage, we obtain the Fourier coefficients by an integration:
\begin{eqnarray}
a_n^{(a,b)} &=& \frac{2}{z_t} \int^{z_t}_{0}{\rm d}z\ \phi^{(a,b)}(\rho_0,z)\ 
	\sin\left(\frac{n \pi z}{z_t} \right) \nonumber \\
&=& \frac2{n \pi} \left[\cos\left(\frac{n \pi a}{z_t} \right) - \cos\left(\frac{n \pi b}{z_t} \right)\right]
	\sin\left(\frac{n \pi z_g}{2 z_t} \right) \frac{2 z_t}{n \pi z_g} \label{eq:design-solution}
\end{eqnarray}
Now we specialize these results to the closed trap with three active electrodes. The trap length is $z_t=2z_0$, the coordinates of the ring and compensators are applied accordingly. The
contributions of the correction electrodes are summarized to one function $\phi^{(c)}$. In order to
analyse the anharmonicity, we expand the potential in a power series around the trap centre
\begin{equation}
\Phi(0,z) = \sum_{j\ {\rm even}} C_j \left(\frac {z-z_0} d\right)^j. \label{eq:taylor}
\end{equation}
There are only even terms, because $z=z_0$ defines a symmetry plane of the trap.
Following the idea of equation (\ref{eq:design-superpos}), we decompose the $C_j$ coefficients into
the contributions of the ring electrode and the compensators,
$C_j =U_0 e_j + U_c d_j$. Defining the tuning ratio $T=U_c/U_0$, we can write 
$C_j = U_0 (e_j + T d_j) = U_0 c_j$. 
%
Here, $e_j$ and $d_j$ are the Taylor coefficients of the sine function times the respective Fourier
coefficients of the particular electrodes, calculated in equation (\ref{eq:design-solution}). They only
depend on the trap dimensions. We search for a geometry where all orders higher than the second
in the expansion (\ref{eq:taylor}) vanish. This results in an over-determined non-linear system of
equations for the variables
$T,\,z_0,\,z_r,\,z_g,\,\rho_0$. The compensator length follows from the other dimensions.
The gap width is constrained by electrical requirements and apart from that, the electrostatic problem is
invariant with respect to rescaling the entire trap. For this reason we leave the gap width and at this point
also the trap length untouched and restrict ourselves to three free parameters: The tuning ratio, the ring
length, and the radius. The optimal tuning ratio is determined by the requirement of a vanishing lowest-order
anharmonicity, $T=-e_4/d_4$. In the experiment, when the geometry is fixed, the electrical tuning is the only
way to compensate for the anharmonicity. Therefore, the trap should be orthogonal, which means that the ion
oscillation frequency is independent of the tuning, or equivalently $d_2=0$. 

This condition 
\begin{figure}[h]
\begin{center}
\includegraphics[width=8cm]{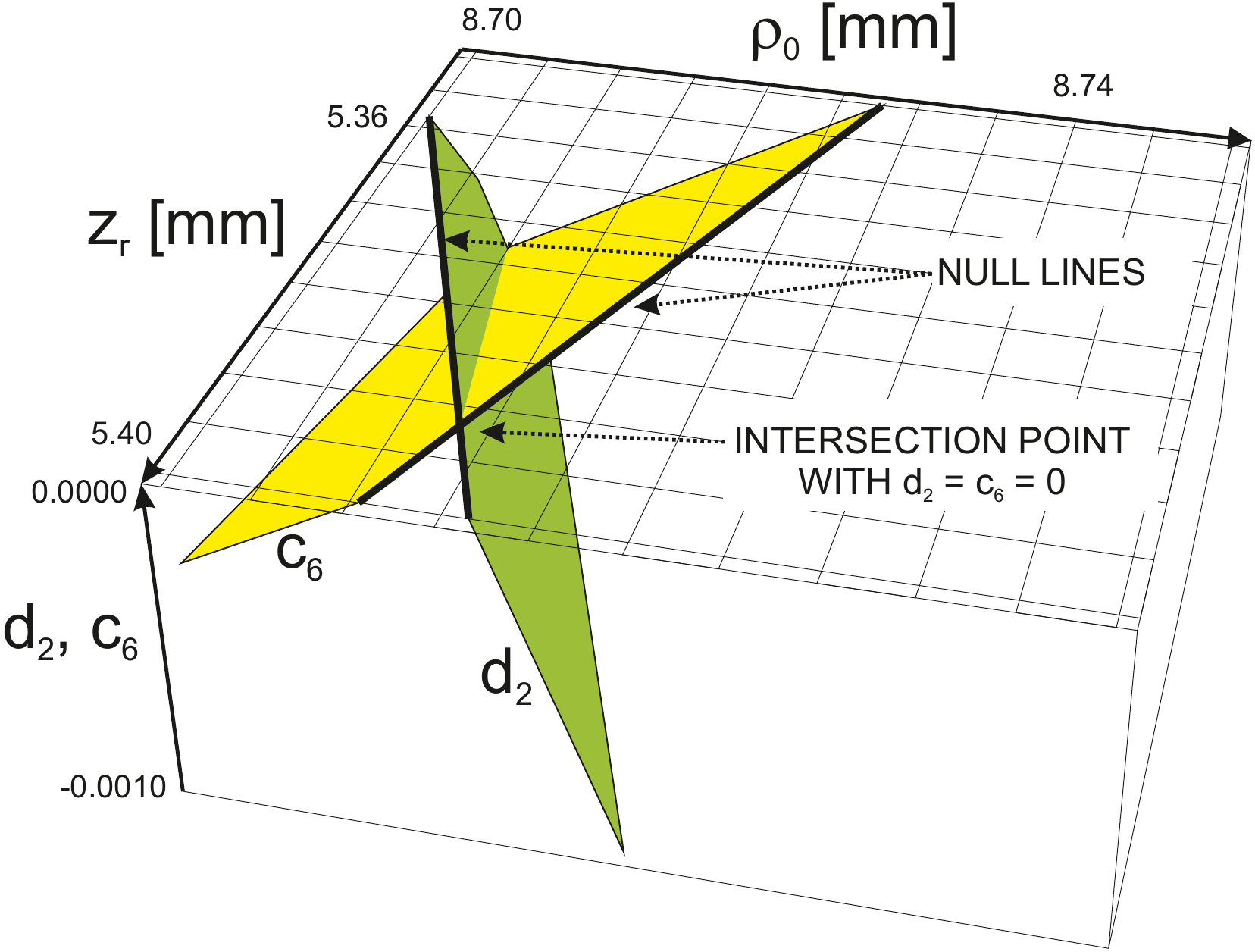}
\caption{$c_6$ (yellow) and $d_2$ (green) as functions of
the ring length and radius. The coordinates of the intersection point, $z_r=5.392$mm and $\rho_0=8.713$mm are the trap dimensions needed for vanishing $c_6$ and $d_2$}
\label{fig:nulllines}
\end{center}
\end{figure}
and the condition for vanishing sixth order of the Taylor expansion are used to find the ring length and radius
with a variational method: we plot $d_2$ and $c_6$ as functions of $z_r$ and $\rho_0$, where $c_6$ is
evaluated with the optimal tuning ratio, see figure \ref{fig:nulllines}. 
Both functions have null lines in
the three-dimensional plot. The lines intersect in the point $(z_r,\rho_0)_{d_2=c_6=0}$. This defines
the optimal geometry, where the 2 lowest-order perturbations can be compensated (brought to zero)
simultaneously without affecting the axial frequency (orthogonal trap). The results are summarized in table \ref{tab:design-results}.

\section{The Half-Open Trap}\label{sec:half-open}
For the present investigations, the cylindrical trap needs to allow efficient light collection on one side, while the other side needs to be open for injection of externally produced ions. At the same time, the trap should maintain the good confinement properties of the closed trap. The first requirement is met by a meshed endcap which will be discussed below. For the trap opening on the opposite side, we use a geometry that approximates the potential calculated for the closed case. This is done by removing one endcap of the closed trap and replacing it by ring electrodes which mimic the effect of this missing endcap. The idea behind this
is to build a mirror image of the closed trap, biased with the opposite voltages -- named the `antitrap'. Then,
for symmetry reasons the potential will be zero
in the plane of the virtual end cap situated between compensator and anti-compensator. The strict application of this idea would duplicate the electrodes and again lead to a
closed trap. However, beyond a specific distance, further modifications are tolerable, because their effect
on the trap centre is shielded by the electrodes in between.

\subsection{Trap Design}
\begin{figure}[h]
\begin{center}
\includegraphics[width=8cm]{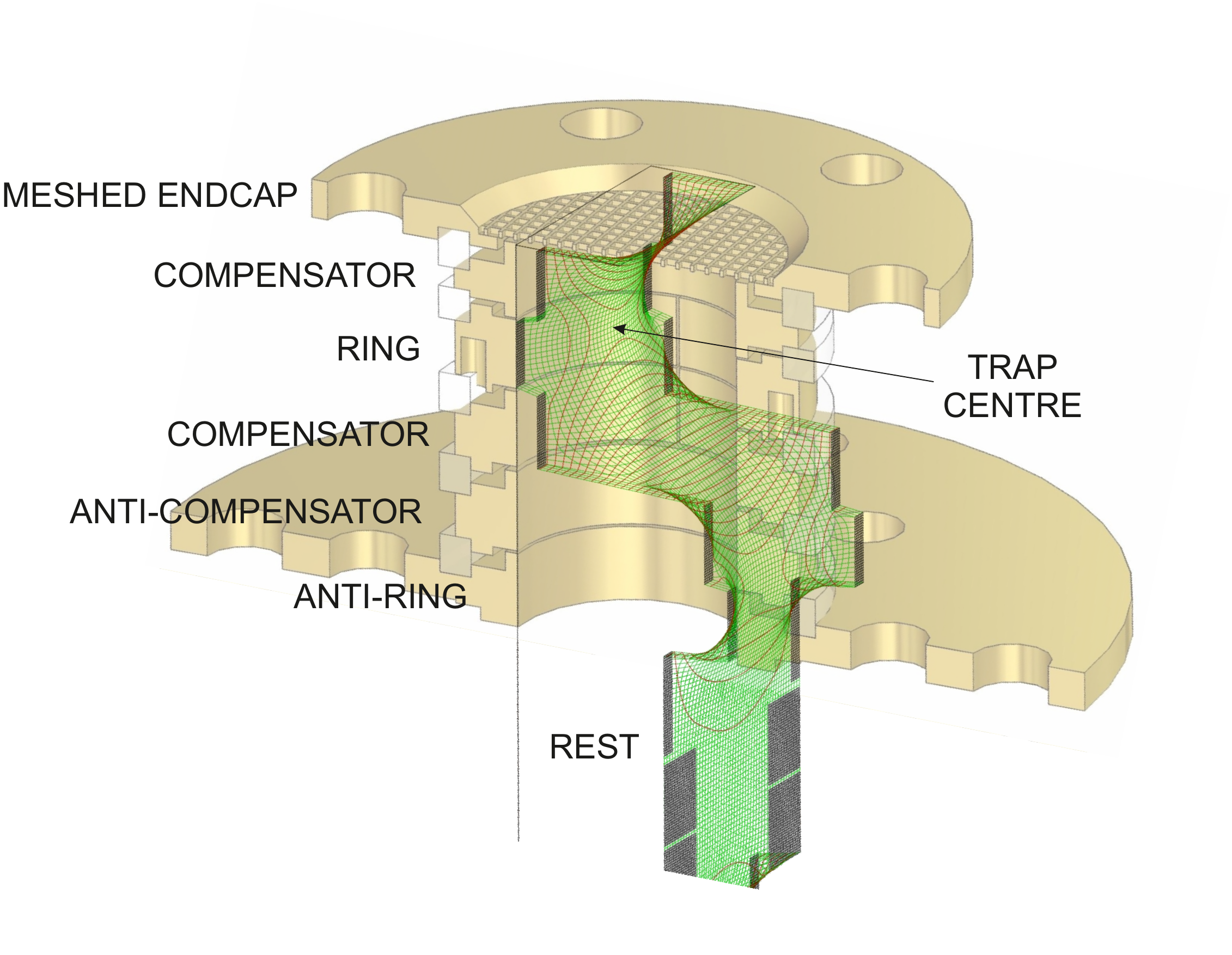}
\caption{Electrode arrangement of our half-open trap (to scale) and the resulting trapping potential. The trap centre (potential minimum) is indicated
\label{fig:design1}}
\end{center}
\end{figure}

Figure \ref{fig:design1} shows a detailed example of the resulting electrode configuration. In order to find the proper geometry and voltage parameters of the half-open trap, we start with the geometry and voltages of the closed trap and
remove the lower end cap to install an anti-compensator, anti-ring and some more electrodes, referred
to as `rest'. This rest is assumed to have a uniform voltage in order to keep the
number of degrees of freedom reasonable. For other purposes, such as ion transport, these electrodes
may be set on different potentials temporarily. The whole assembly is modelled with the length $z_t=8z_0$, an arbitrary choice.
The anti-compensator is set to the negative voltage of the compensator, whereas the voltages $U_{ar}=T_1U_0$
of the anti-ring and $U_{rest}=T_2U_0$ of
the rest are still to be determined by a model fit. Another two free parameters are found in the
anti-compensator and anti-ring length, $z_{ac}$ and $z_{ar}$, respectively. Between the electrodes, we keep
the same gap width as in the closed trap, with one exception: The gap between the lower compensator
electrode and the anti-compensator is doubled, in order to keep the virtual end cap at the original position of the removed real endcap.
As we did for the closed trap, we determine the contribution $\phi^{(i)}(\rho,z)$ of each electrode
in the new arrangement. According to formula (\ref{eq:design-superpos}), the trapping potential
$\Phi(\rho,z)$ is again given as a linear combination of these functions,
multiplied by the respective voltages $U_i$, which have been defined above.
The resulting function is evaluated in the plane of the virtual end cap, $z=z_0$. This is a function
of one variable, the radius, with four fit parameters.
Then, ten equidistant points ($\rho/\rho_0=0,\,1,\,2,\,...,\,9$) in the plane are chosen for an
artificial data set to have the value zero.
A fit routine finds the best values for the model parameters to describe the artificial data. This brings
the potential in the relevant plane as close to zero as possible.

The real trap however will be subject to machining inaccuracies on the level of $\mu$m. This may introduce non-zero odd terms in the Taylor expansion (\ref{eq:taylor}), which shift the trap centre. To account for this, we round the calculated dimensions on that level and fix them to values listed in table
\ref{tab:design-results}. With these definitions we can find an estimation for realistic expansion
coefficients (see table \ref{tab:design-coeff}). 
\begin{table}[h]
\begin{center}
\caption{
Results of the semi-analytical calculations for the closed trap and of the fit method for the half-open trap. Lengths and radius in mm, tunings are dimensionless
\label{tab:design-results}}
\begin{tabular}{cccccccccc}\\
\hline\hline
  \raisebox{0pt}[12pt][6pt]{$z_0$} &	$z_r$ &	$\rho_0$ &	$z_c$ &	$T$
	&	$z_g$ &	$z_{ac}$&	$z_{ar}$&	$T_1$ &		$T_2$\\
\hline
  \raisebox{0pt}[12pt][6pt]{9.000} &	5.392 &	8.713 &	5.904 &	0.799404
	&	0.200 &	6.200 &	5.740 &	-1.02071 &	-0.402493\\
\hline\hline
\end{tabular}
\end{center}
\end{table}
\begin{table}[h]
\begin{center}
\caption{
Coefficients for the Taylor expansion (\ref{eq:taylor}) of the potential in the half-open trap
\label{tab:design-coeff}}
\begin{tabular}{cccccc}\\
\hline\hline
  \raisebox{0pt}[12pt][6pt]{$c_1$}	& $c_2$	& $c_3$
	& $c_4$	& $c_5$	& $c_6$\\
\hline
  \raisebox{0pt}[12pt][6pt]{$-1.6\cdot10^{-5}$}	& $-0.522644$	& $-1.8\cdot10^{-5}$
	& $-1.4\cdot10^{-5}$		& $-1.1\cdot10^{-5}$		& $-1.8\cdot10^{-5}$\\
\hline\hline
\end{tabular}
\end{center}
\end{table}
Now we can evaluate the perturbative expressions for
the shift of the axial frequency due to non-zero 4$^{th}$ and 6$^{th}$ order contributions by use of equations (\ref{eq:design-p4}) and
(\ref{eq:design-p6}), which yields $\Delta\omega_z/\omega_z \approx 8.0 \cdot 10^{-10}$ for Ar$^{13+}$ ions with an
energy corresponding to {60}\,K in the axial oscillation, if the trap voltage is {-10}\,{V}.

The real trap as characterized by the rounded coefficients is also no longer orthogonal, which is quantified by
$d_2=4.5 \cdot 10^{-5}$. This has the effect that the axial frequency is shifted with the correction
voltage according to $\partial \omega_z/\partial U_c \approx 2\pi\cdot {1.6}$\,{Hz/V}. Keeping
in mind that the frequency itself is around $2\pi \cdot {0.37}$\,{MHz} and that the correction
voltage is stable in the sub-mV range, this translates to a ppb effect.

In our specific realization, the mesh in the upper endcap (spectroscopy mesh) is made of a copper foil with 5.08\,{$\mu$m} thickness. It has a density of
2.07 wires per mm with 96.5\,{$\mu$m} width. In contrast to the solid electrodes, the mesh
has been plated with a silver layer of only 5\,{$\mu$m} and again less than a micrometer of
gold. These characteristics correspond to 60\% light transmission and an areal mass density of
60\,g/m$^2$.

Since the mesh serves as upper end cap of the spectroscopy trap, small deformations
of this very fine piece of metal would lead to electrostatic potential distortions and corrupt
precision. We estimate the gravitational and electrical stress acting on the
mesh to be 0.6\,Pa and below 7\,mPa,
respectively. The latter is calculated with the maximum electrical field
${\cal {E}}=40$\,{kV/m} and the analogy of a capacitor: The two plates experience
an areal force density of $p_{\mathcal E}=\varepsilon_0/2\ |{\cal{E}}|^2$, if there is the electrical
field $\mathcal E$ between them.
However, this contribution is negligible compared to the gravitational effect, which in itself has not been observed in the experiment.

\subsection{Light Collection Efficiency} \label{sec:design-mesh}
With the half-open design in which the upper endcap of a closed trap is replaced by a
transparent mesh we obtain a solid angle of {1.77}\,{sr}, which is {14}\,{\%} of the full sphere, for light collection, as compared to about 2\,\% of the full sphere in the case of the open-endcap trap, see above. The fraction
of the light power arriving at each end face is close to {20}\,{\%} of the total emitted power due to the angular intensity distribution (\ref{eq:distr}).
\begin{figure}[h]
\begin{center}
\includegraphics[width=6cm]{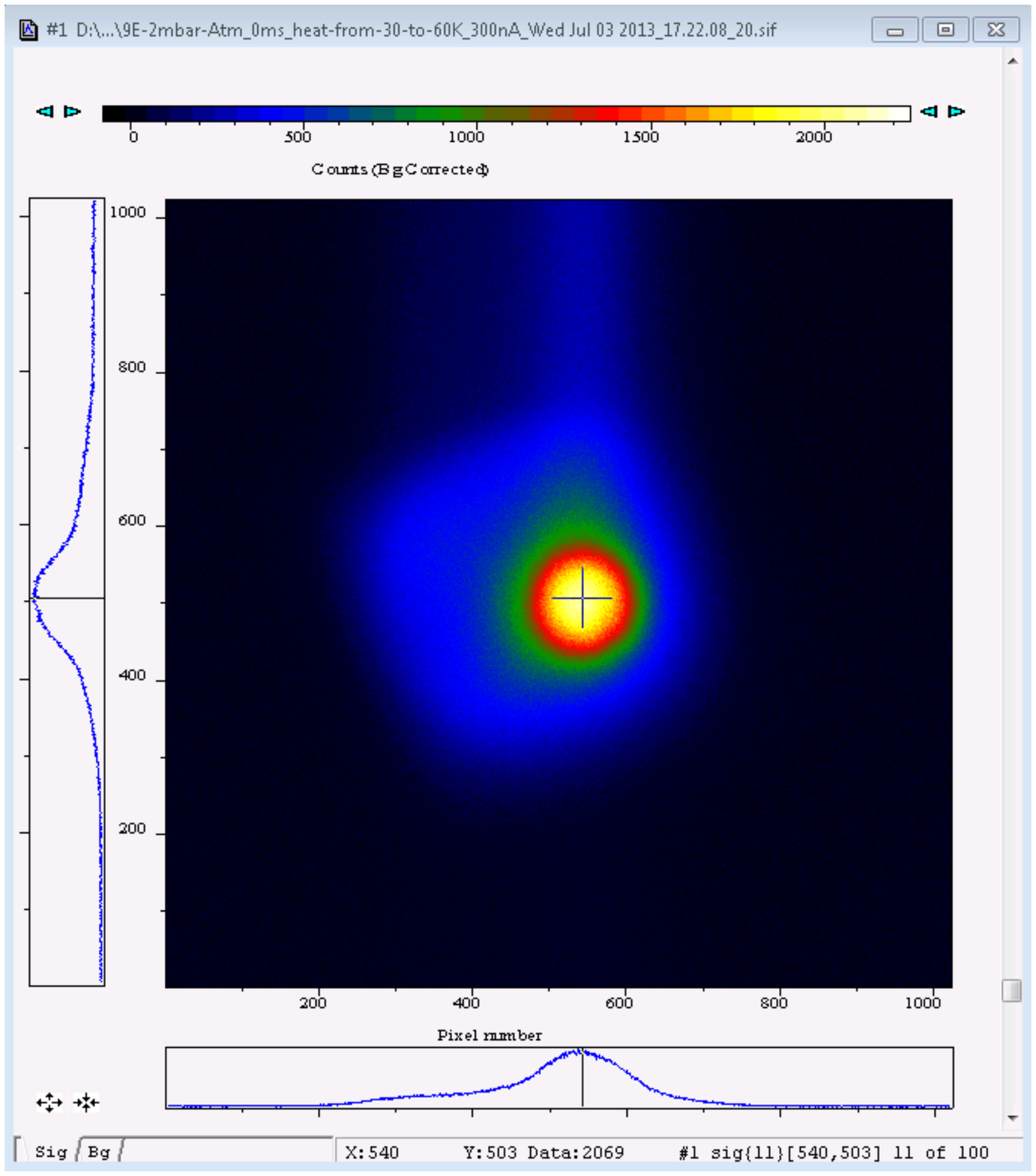}
\caption{CCD image of fluorescing ions taken from above through the optical mesh. For details see text
\label{image}}
\end{center}
\end{figure}
To illustrate light collection, figure \ref{image} shows a false-colour CCD image of a fluorescing ion cloud confined in the trap as seen from above through the mesh. To that end, the light is transferred into a 4\,mm by 4\,mm (176 by 176 pixel) image guide which relays the light from trap centre with an efficiency of about 50\% to a sensitive CCD camera (ANDOR IKON-N) with 1024 by 1024 pixel resolution. The fluorescence in this case is due to collisions of residual gas and argon with an electron beam of several nA current and keV energy reflected back and forth through the charge breeding section of the trap, in close similarity to the mini-EBIS described in \cite{alonso}. 

In the next step of the experiment, Ar$^{13+}$ ions will specifically be charge-bred from argon gas using a 1855\,eV electron beam and confined in the trap. With these ions, laser-microwave double resonance spectroscopy will be performed as described in detail in \cite{pra1,zeeman}. This is also an excellent test case for further studies with heavier ions of higher charge states as available from the HITRAP \cite{kluge,fra} facility in its present configuration and future configurations in the framework of the FLAIR project.  

\section{Conclusion}
In a half-open cylindrical Penning trap as described above, light collection is about one order of magnitude more efficient than in the 'classical' open-endcap design. We have demonstrated fluorescence light detection from such a trap within the ARTEMIS experiment using unspecific fluorescence emission from confined ions produced by and collided with an electron beam traversing the trap. This is a necessary prerequisite for the production and confinement of Ar$^{13+}$ which are next to be examined by high-precision laser-microwave double-resonance spectroscopy. Following this, externally produced heavier ions from the HITRAP facility will be captured, confined and studied by the same methods. This is a unique method for high-precision determinations of the magnetic moments of both the ionic nucleus and the bound electron(s) and stringent tests of corresponding models within QED of bound states and diamagnetic shielding.
\section{Acknowledgements}
This work has been funded by the DFG under grant number BI647/4-1 and the International Max Planck Research School for Quantum Dynamics.


\begin{thebibliography}{99}
\bibitem{MAJ04} Major, F.G., Gheorghe, V.N. and Werth, G.: \textit{Charged Particle Traps}, Springer (2004)
\bibitem{werth09} Werth, G., Gheorghe, V.N. and Major, F.G. : \textit{Charged Particle Traps II}, Springer (2009)
\bibitem{gho} Ghosh, P.K.: \textit{Ion Traps}, International series of monographs in physics, Oxford (1995)
\bibitem{gab89} Gabrielse, G., Haarsma, S., Rolston, S.L.: Int. J. Mass Spectr. Ion Proc. {\bf 88} 319 (1989) 
\bibitem{bhar} Bharadia, S., Vogel, M., Segal, D.M. and Thompson, R.C.: Appl. Phys. B {\bf 107} 1105 (2012) 
\bibitem{spec} Vogel, M., Winters, D., Segal, D.M. and Thompson, R.C.: Rev. Sci. Instr. {\bf 76} 103102 (2005) 
\bibitem{spec2} Andjelkovic, Z., Cazan, R., N\"ortersh\"auser, W., Bharadia, S., Segal, D.M., Thompson, R.C. and Vogel, M.: Phys. Rev. A {\bf 87} 033423 (2013)
\bibitem{haff} H\"affner, H. et al.: Phys. Rev. Lett. {\bf 85} 5308 (2000)
\bibitem{verd} Verd\'{u}, J. et al.: Phys. Rev. Lett. {\bf 92} 093002 (2004)  
\bibitem{sturmSi} Sturm, S. et al.: Phys. Rev. Lett. \textbf{107} 023002 (2011)
\bibitem{wag} Wagner, A. et al.: Phys. Rev. Lett. {\bf 110} 033003 (2013)
\bibitem{pla} Stahl, S., Galve, F., Alonso, J., Djekic, S., Quint, W., Valenzuela, T., Verdu, J., Vogel, M. and Werth, G.: Eur. Phys. J. D {\bf 32} 139 (2005) 
\bibitem{pla2} Castrejon-Pita, J.R. and Thompson, R.C.: Phys. Rev. A {\bf 72} 013405 (2005)
\bibitem{gal} Galve, F., Fernandez, P. and Werth, G.: Eur. Phys. J. D {\bf 40} 201 (2006)
\bibitem{bush} Bushev, P., Stahl, S., Natali, R., Marx, G., Stachowska, E., Werth, G., Hellwig, M. and Schmidt-Kaler, F.: Eur. Phys. J. D {\bf 50} 97 (2008)
\bibitem{gabbash} Goldmann, J. and Gabrielse, G.:Phys. Rev. A {\bf 81} 052335 (2010)
\bibitem{cast} Castrejon-Pita, J.R. et al.: J. Mod. Opt. {\bf 11} 1581 (2007)
\bibitem{wine} Maiwald, R., Leibfried, D., Britton, J., Bergquist, J.C., Leuchs, G. and Wineland, D.J.: Nature Physics {\bf 5} 551 (2009)
\bibitem{a1} Yu, N., Nagourney, W. and Dehmelt, H.: J. Appl. Phys. {\bf 69} 3779 (1991)
\bibitem{a2} Schrama, C., Peik, E., Smith, W. and Walther, H.: Opt. Commun.
{\bf 101} 32 (1993)
\bibitem{a3} Deslauriers, L. et al.: Phys. Rev. Lett. {\bf 97} 103007 (2006)
\bibitem{pra1} Quint, W., Moskovkin, D., Shabaev, V.M. and Vogel, M.:
Phys. Rev. A {\bf 78} 032517 (2008)
\bibitem{dr} von Lindenfels, D. Brantjes, N., Birkl, G., Quint, W., Shabaev, V.M. and Vogel, M.: Can. J. Phys. {\bf 89} 79 (2011) 
\bibitem{zeeman} von Lindenfels, D., Wiesel, M., Quint, W., Glazov, D., Shabaev, V.M., Birkl, G. and Vogel, M.: Phys. Rev. A {\bf 87} 023412 (2013)
\bibitem{pr} Vogel, M. and Quint, W.: Physics Reports {\bf 490} 1-47 (2010)
\bibitem{gab} Gabrielse, G. and Mackintosh, F.C.: Int. J. of Mass Spectrom. Ion Processes, {\bf 57}(1), 1984.
\bibitem{han} Hanneke, D. et al.: Phys. Rev. Lett. {\bf 100} 120801 (2008)
\bibitem{sensors} Vogel, M., Quint, W. and N\"ortersh\"auser, W.: Sensors, {\bf 10} (2169), 2010.
\bibitem{alonso} Alonso, J. et al.: Rev. Sci. Inst. \textbf{77}, 03A901 (2006)
\bibitem{kluge} Kluge, H.-J. et al., Advances in Quantum Chemistry {\bf 53}, 83 (2007).
\bibitem{fra} Herfurth, F. et al.: Int. J. Mass Spectr. {\bf 251}, 266 (2006).

\end{thebibliography}
\end{document}